\documentclass{ws-ijmpa}

\usepackage[super]{cite}
\usepackage{xcolor}
\usepackage[verbose,hypertexnames=false]{hyperref}
\hypersetup{colorlinks=false,allbordercolors=blue,pdfborderstyle={/S/U/W 1}}

\usepackage{pdflscape}
\usepackage{amsmath}
\usepackage{amsfonts}
\usepackage{amssymb}
\usepackage{graphicx}
\usepackage[titletoc]{appendix}
\usepackage{color}
\usepackage{hyperref}
\usepackage{cleveref}
\usepackage[rightcaption]{sidecap}
\usepackage{comment}
\usepackage{soul}
\usepackage{cancel}

\usepackage{dcolumn}

\usepackage{longtable}
\usepackage{floatrow}
\usepackage{array}
\usepackage{ctable}
\usepackage{multirow}
\usepackage{siunitx}
\usepackage{tabularx}
\usepackage{booktabs}
\usepackage{supertabular}

\graphicspath{{Graphics/}}

\def\be{\begin{equation}}
\def\ee{\end{equation}}
\def\bea{\begin{eqnarray}}
\def\eea{\end{eqnarray}}

\definecolor{vividviolet}{rgb}{0.62, 0.0, 1.0}
\definecolor{amaranth}{rgb}{0.9, 0.17, 0.31}
\definecolor{palatinateblue}{rgb}{0.15, 0.23, 0.89}
\definecolor{brightpink}{rgb}{1.0, 0.0, 0.5}
\definecolor{cornflowerblue}{rgb}{0.39, 0.58, 0.93}
\definecolor{deepcarminepink}{rgb}{0.94, 0.19, 0.22}
\definecolor{radicalred}{rgb}{1.0, 0.21, 0.37}

\hypersetup{ linktoc=all,
    colorlinks, linkcolor={palatinateblue},
    citecolor={brightpink}, urlcolor={amaranth}
}

\begin{document}

\markboth{F. Lottatori}{Parametrizing superfluid dark matter with rational approximations}

%
\catchline{}{}{}{}{}
%

\title{Parametrizing superfluid dark matter with rational approximations}

\author{Francesco Lottatori}
 
\address{Department of Physics “E.R. Caianiello”, University of Salerno, Via Giovanni Paolo II 132, 84084 Fisciano (SA), Italy.\\
flottatori@unisa.it}

\maketitle


\begin{abstract}

We investigate how a spatially modulated real scalar background $\phi(\vec{x})$ can modify phonon propagation in the context of Superfluid dark matter (SFDM). Using a simple toy model with quartic condensate and coupling $-g\phi^2|\Psi|^2$, we derive the local equation of state and the effective sound velocity $c_s(\vec{x})$. For $g>0$, modulation tends to increase the effective mass of the condensate and make the medium less rigid, suppressing $c_s^2\propto m_{\Psi,\mathrm{eff}}^{-4}$ up to a ``dust-like'' regime, $c_s^2\to 0$. We implement this modulation for the background scalar field by imposing rational profiles, through Pad\'e radial profiles, and show the corresponding variation of $c_s^2(r)$ for different $g$, discussing implications for the structure of SFDM cores and the possible formation of inhomogeneous regions of dark matter. 

\end{abstract}



\section{Introduction}

Cosmological observations  indicate that most of the matter in the Universe is non-baryonic and non-relativistic on large scales\cite{Planck:2018vyg} and a cosmological constant problem is still persistent within the so-called standard cosmological model \cite{Belfiglio:2022qai,Belfiglio:2023rxb}. In particular, on galactic scales, important regularities are empirically observed that connect the distributions of baryonic matter and dynamics with low intrinsic dispersion \cite{PhysRevLett.117.201101}. In this context, MOdified Newtonian Dynamics (MOND) may describe the low acceleration regime without invoking modification of Einstein's gravity \cite{1983ApJ...270..365M,1984ApJ...286....7B}. The effective model, however, faces considerable difficulties when attempting to describe large-scale dynamics, particularly with galaxy clusters. SFDM  alleviates the MOND inconsistencies by postulating a superfluid phase in galactic cores and cold dark matter behaviour, as imposed by the standard cosmological model, on cosmological scales \cite{Berezhiani:2015bqa}. For a recent comprehensive review of the theoretical foundations, galactic phenomenology and connection with MOND in superfluid dark matter, see Ref.~\cite{Berezhiani:2025sfdm}.

According to this new perspective, SFDM offers an interpretation of dark matter in which the microphysics of the dark sector plays a significant role on galactic scales. The superfluid phase can systematically help to alleviate certain difficulties that arise on small scales within the cold dark matter paradigm, such as the presence of galactic cores, the regularity of baryonic scaling relations, and the empirical link between baryonic acceleration and total acceleration, whilst preserving the behaviour described by the standard cosmological model $\Lambda$CDM on large scales.

In the superfluidity regime, the low-energy degrees of freedom are phonons, i.e., Goldstone bosons associated with the spontaneous breaking of a global $U(1)$ symmetry, which can mediate an additional force by coupling to baryons \cite{Berezhiani:2015bqa}. 

A further perspective may be gained from emerging holographic approaches to the dark sector. In particular, holographic descriptions based on geometric invariants have been proposed as possible unified scenarios for dark matter and dark energy \cite{Aviles:2011ak}. The SFDM model used here does not stem from the proposals mentioned above; however, it leaves the door open to considering the possible existence of a scalar background $\phi$, interpretable as an effective degree of freedom, capable of encoding both environmental and geometric properties of the system under analysis, and of locally modifying the phononic response of the SFDM condensate.

In this context, we may wonder how the presence of phonons may affect the background. To do so, we study the effect of a real scalar field $\phi$ with a static but spatially modulated profile on the sound speed in the medium. This may emulate a modification of Einstein's gravity with effective couplings, often explored in the literature, see e.g. \cite{Abedi:2018lkr}.

The effective reconstruction is modeled by rational approximations, physically supported by the fact that any SFDM fluid may reconstruct dark matter along galaxies and, therefore, its influence has to be significant in specific regions, turning into zero, or into a constant value, as the galaxy ends. This preliminary analysis opens avenues to compare these environmental modulations with gravitational-lensing constraints and distinguish SFDM from other paradigms, as well as testing the viability of spatially varying effective properties in the superfluid core.

\section{Modelling SFDM and its sound speed}

Let us consider a complex scalar field $\Psi$ representing the dark matter condensate, modelled according to the SFDM approach, and a real scalar field $\phi$ as the background for this model, providing
\begin{equation}
\mathcal{L}=
\partial_\mu\Psi^*\partial^\mu\Psi
-m^2|\Psi|^2
-\frac{\lambda}{2}|\Psi|^4
+\frac{1}{2}\partial_\mu\phi\,\partial^\mu\phi
-\frac{1}{2}m_\phi^2\phi^2
-g\,\phi^2|\Psi|^2\,\,\,,
\label{eq:1}
\end{equation}
with $\lambda>0$ for the stability of the minimal model, $g$ the coupling constant considered dimensionless for reasons of renormalizability of the theory, $m$ the mass of the constituents of the dark matter condensate, and $m_\phi$ the mass of the background scalar field. The background $\phi(\vec{x})$ modulates the effective mass of the SFDM condensate\footnote{With the constraint $m_{\Psi,\mathrm{eff}}^2>0$ if $g<0$, to maintain the essential mechanism of spontaneous symmetry breaking that the system undergoes.}:
\begin{equation}
m_{\Psi,\mathrm{eff}}^2(\vec{x})=m^2+g\,\phi(\vec{x})^2\,\,\,.
\label{eq:2}
\end{equation}
The dynamics of the scalar field $\phi$ in the presence of the condensate also induces a substantial change in the effective mass of the latter, following the relationship below:
\begin{equation}
m_{\phi,\mathrm{eff}}^2(\vec{x})=m_\phi^2+2g|\Psi|^2 \,\,\,.
\label{eq:3}
\end{equation}
The resulting equations of motion are given by:
\begin{equation}
\Box\Psi+m_{\mathrm{eff}}^2(\phi)\Psi+\lambda|\Psi|^2\Psi=0,\qquad \Box\phi+m_{\phi,\mathrm{eff}}^2\phi=0.
    \label{eq:4}
\end{equation}

\subsection{Singling out the scenarios}

We now suppose a viable static, spatially modulated profile for the field $\phi(\vec{x})$, isolating the direct effect on the phonon sector, in line with what one  expects for spiral galaxies \cite{PhysRevLett.117.201101}. In the non-relativistic limit and at leading order in gradients, a quartic condensate is well described by the equation of state:
\begin{equation}
P=\frac{\lambda}{8m^4}\,\rho^2\,\,\,,
\label{eq:5}
\end{equation}
where $\rho$ is the mass density.
The most obvious way to include the background is the substitution $m\to m_{\Psi,\mathrm{eff}}(\vec{x})$, which gives us the following new equation of state, this time local:
\begin{equation}
P(\vec{x})=\frac{\lambda}{8\,m_{\Psi,\mathrm{eff}}(\vec{x})^4}\,\rho(\vec{x})^2\,\,\,.
\label{eq:6}
\end{equation}
From $c_s^2=\partial P/\partial\rho$, at a fixed position, we obtain the relationship for the local speed of sound in the medium:
\begin{equation}
c_s^2(\vec{x})=\frac{\lambda}{4\,m^4_{\Psi,\mathrm{eff}}(\vec{x})}\,\rho(\vec{x})\,\,\,.
\label{eq:7}
\end{equation}
Therefore, an increase in the field $\phi$ implies an increase in the effective mass $m_{\Psi,\mathrm{eff}}$, thus modulating the speed of sound $c_s$, which decreases as it is algebraically suppressed, making the medium less rigid. Specifically, the non-uniformity of $\phi(\vec{x})$ also causes $c_s$ to acquire a spatial dependence, reducing the local Jeans length to 
\begin{equation}
\lambda_J(\vec{x}) = \frac{c_s(\vec{x})}{\sqrt{G\,\rho(\vec{x})}}\,\,\,,
\label{eq:8}
\end{equation}
favouring large-scale clustering in regions where the speed of sound is lower, increasingly approaching, for $g>0$, a ``dust-like" behaviour in the limit $c_s^2\to 0$. Obviously, the dynamics remain stable as long as the compressibility of the medium remains physical, i.e. $\partial P/\partial\rho>0$, or similarly $c_s^2\ge 0$, and as long as $m_{\Psi,\mathrm{eff}}^2>0$.

\section{Pad\'e profile for $\phi(r)$ and sound speed suppression}

From here on, we assume spherical symmetry and static profiles, so $\vec{x}\to r$ and $c_s(\vec{x})\to c_s(r)$. In particular, we will also use $x\equiv r/R_{\rm SF}$ as a dimensionless variable in the Pad\'e profile.

To model a smooth radial profile between a central value and a possible external plateau for the background field, we use the so-called ``Pad\'e approximant'', as it is more flexible for such purposes than other methodologies, such as Taylor series, e.g. \cite{Basdevant1972Pade}.

In cosmology, Padé approximants were first introduced as a rational alternative to Taylor expansions of cosmographic observables as functions of redshift in Ref.~\cite{Gruber:2013wxa}, owing to their remarkable convergence properties. In this work, we draw inspiration from this original idea, but apply it in a different context, namely to the phenomenological radial ansatz for the external scalar field $\phi(r)$, with the aim of obtaining a smooth profile with controlled asymptotic behaviour for the latter.

Thus, setting the same order of numerator and denominator, plus the simplest case, namely the lowest nontrivial order in numerator and denominator, a Pad\'e approximant of degree $(1,1)$ for $\phi$ reads
\begin{equation}
    \phi(r)=\phi_{in}\frac{1+\alpha \Bigl(\frac{r}{R_{SF}}\Bigr)}{1+\beta \Bigl(\frac{r}{R_{SF}}\Bigr)}\,\,\,,\quad x\equiv\frac{r}{R_{\rm SF}}\in[0,1]
    \label{eq:9}
\end{equation}
which describes a growing profile from $\phi_{\rm in}=\phi(0)$ towards an asymptotic value $\phi_{\infty}=\lim_{r\to\infty}\phi(r)=\phi_{\rm in}\,\frac{\alpha}{\beta}$, all with respect to the characteristic scale $R_{SF}$, the radius of the superfluid dark matter halo that would surround the spiral galaxies under consideration.

We therefore use this technique to generate a spatial ansatz for the field $\phi$, in order to interpolate the asymptotic zones we need for the latter, linked to the phase transition that must result for the system and so as not to alter the mechanism of spontaneous breaking of the $U(1)$ symmetry analysed with $g>0$. In this respect, we also analyse the case with $g<0$ to obtain a more detailed picture.

Now, to find the relationship between $c_s^2$ and $g$ and $\phi(r)$, simply combine Eqs.~\eqref{eq:2} and \eqref{eq:7} to obtain
\begin{equation}
c_s^2(r)=\frac{\lambda}{4}\,
\frac{\rho(r)}{\Bigl(m^2+g\,\phi(r)^2\Bigr)^2}\,\,\,,
\label{eq:10}
\end{equation}
which tends towards zero for increasing values of $g>0$ and of the background field, which we recall is described by the previously chosen Pad\'e profile.

\begin{figure}[ht]
\centering

\begin{minipage}[t]{0.51\linewidth}
  \centering
  \includegraphics[width=\linewidth]{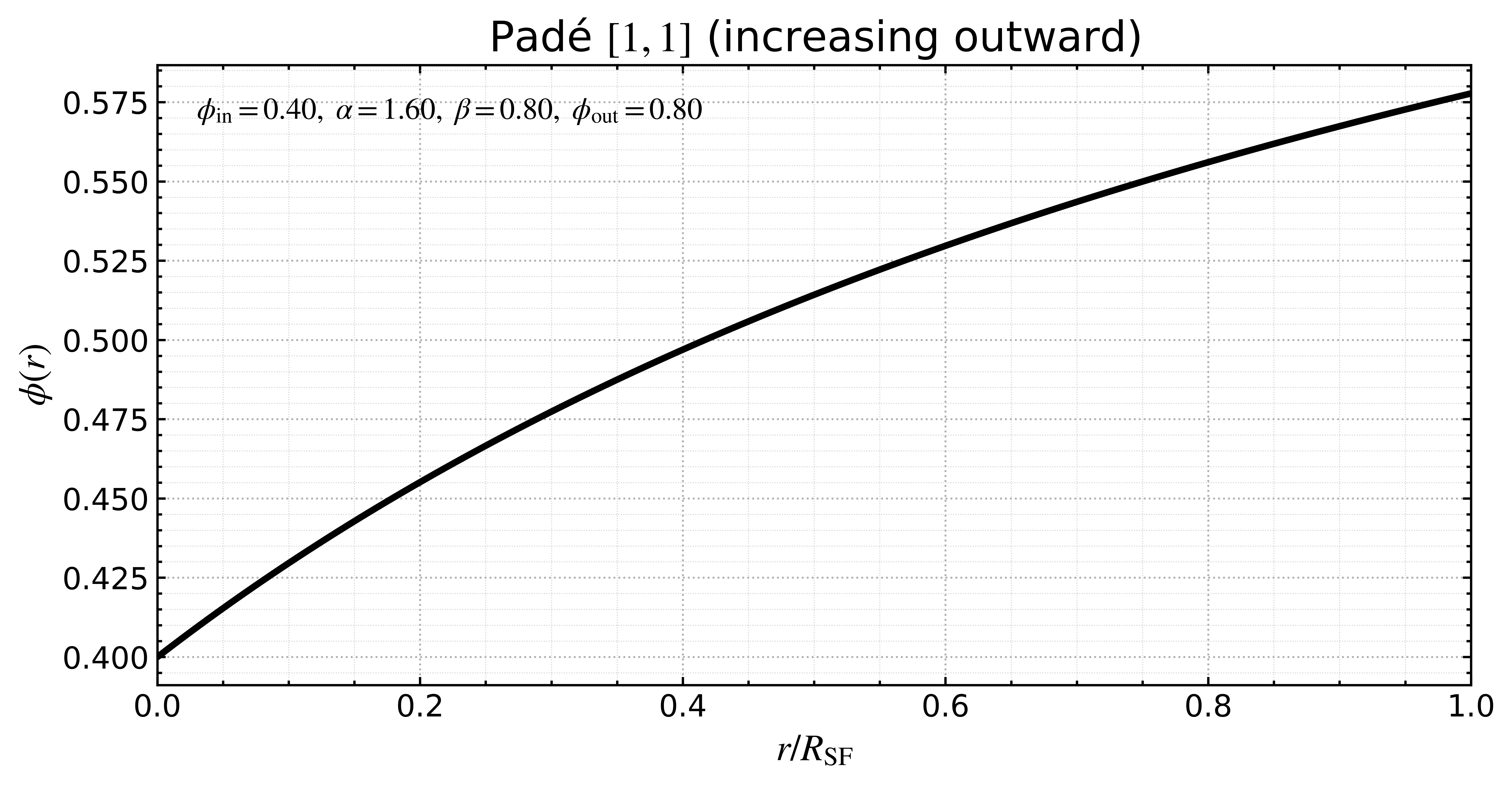}
\end{minipage}
\hfill
\begin{minipage}[t]{0.48\linewidth}
  \centering
  \includegraphics[width=\linewidth]{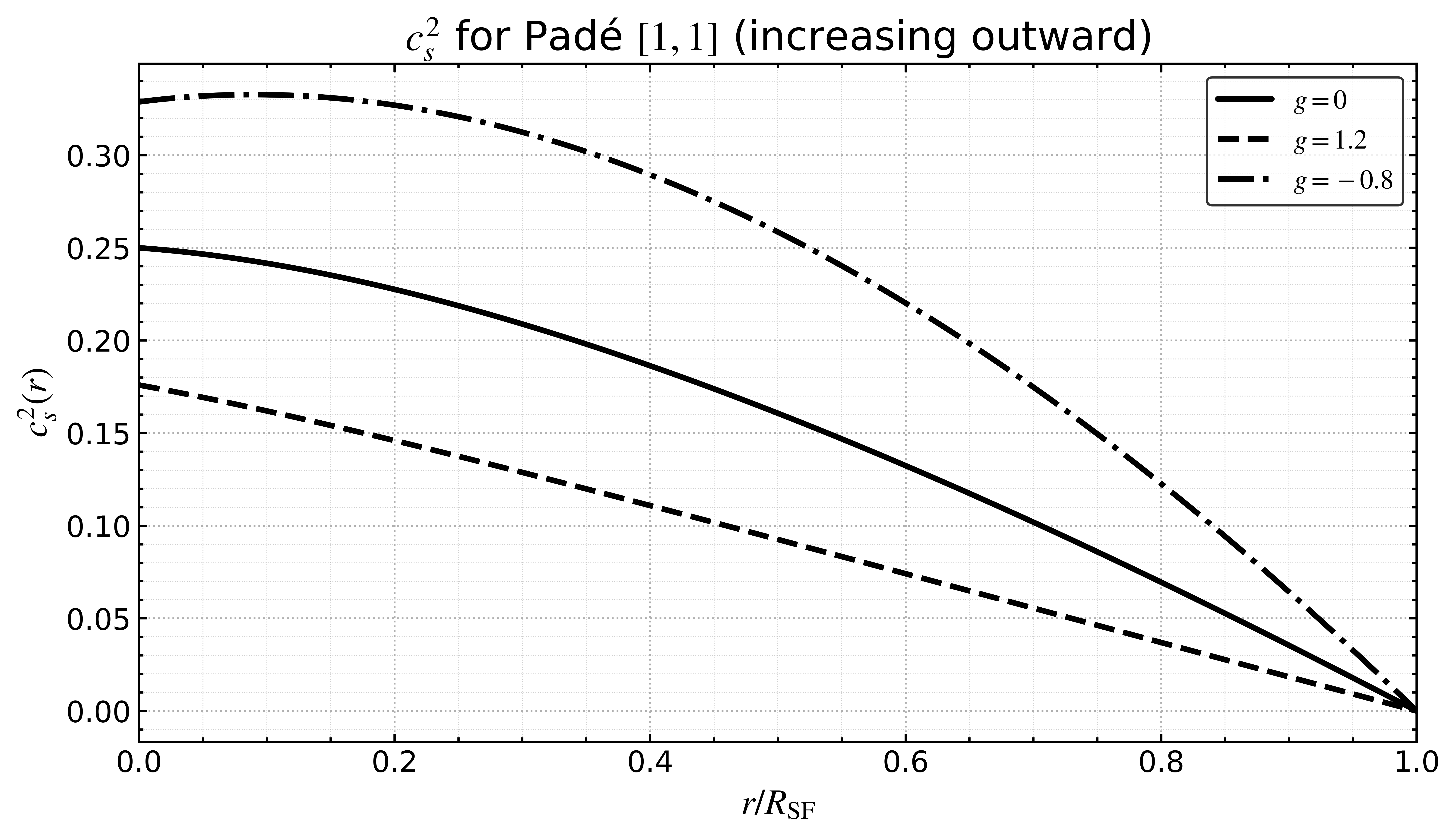}
\end{minipage}

\caption{Left: increasing Pad\'e profile for the scalar background $\phi(r)$, showing a smooth interpolation between the central value and the outer plateau on the scale $R_{SF}$. Right: profile of $c_s^2(r)$ with increasing Pad\'e background for different values of $g$; as $g>0$ and $\phi(r)$ increase, the suppression becomes stronger up to regions where $c_s^2 \rightarrow 0$. The graph legends show the values of the parameters adopted as benchmarks.}
\label{fig:pade_profiles}

\end{figure}

In Fig.~\ref{fig:pade_profiles}, the Pad\'e profile used as a function of distance and the values of the speed of sound, again as a function of distance, are illustrated for different values of the coupling constant $g$, emphasising in particular the difference between a positive and negative value for the latter.

It is noteworthy, and rightly so, that acoustic instability occurs for negative values of the coupling constant $g$ of the interaction introduced between the SFDM condensate field and the background field: in fact, in this situation, this interaction would be repulsive between the fields under consideration, thus suppressing perturbations with a consequent decrease in clustering events. Let us focus now on the functional trends for sound speed that we obtain in the SFDM core with the addition of the aforementioned coupling with the introduced scalar field: if we consider the well-known gravitational lensing effect, we find inhomogeneous dark matter bubbles, which is currently unjustified, since if dark matter were to interact weakly and only gravitationally, as is currently believed, the dominant symmetry should be the central one. Furthermore, it is typically observed that the brightest regions, i.e. those composed of a greater number of baryons, also have more dark matter\footnote{In this regard, there is one important exception to note, namely dwarf galaxies.}.

\section{Conclusion}

We started from the minimal SFDM model with a quartic condensate in its associated Lagrangian density,  introducing a self-coupling $g\phi^2|\Psi|^2$. We formulated toy models based on static Pad\'e-type profiles, resulting in a spatially dependent effective mass. We thus explored the corresponding sound speeds and speculate that these scenarios may be adapted to rotation curves of spirals. 

The findings presented in this paper constitute an initial preliminary analysis in which Padé rational profiles are used to parameterise a scalar background introduced for the purpose of studying the local suppression of the speed of sound in an SFDM condensate. This paper should therefore be regarded as a first step towards a more extensive study, in which the radial profile and the scalar coupling will be constrained by galactic observables.

The model described in the previous sections may prove to be a pioneering approach to explaining some of the open questions in cosmology, particularly on small scales. Indeed, a spatially dependent speed of sound can alter the effective Jeans scale and the clustering properties of dark matter. Furthermore, the minimal SFDM scenario, on which this work is based, offers a possible alternative interpretation of dark matter; however, it requires a much more structured dynamical analysis before quantitative conclusions can be drawn regarding cosmological tensions.

In future studies, more physical Padé series will be explored. Applications to galaxies will then be studied and we will investigate the consequence of SFDM on dark energy, checking how dynamic dark energy influences dark matter, trying to obtain a unified dark energy-dark matter model \cite{Dunsby:2016lkw}. From this perspective, it will also be interesting to investigate possible links with holographic or emergent descriptions of the dark sector, such as those proposed in Ref.~\cite{Aviles:2011ak}. Last but not least, we will check observational signatures and compare our paradigm with the most established dark matter models, e.g. axions, WIMPs, etc., and with possible alternatives \cite{Luongo:2025iqq}.

\section*{Acknowledgements}
It is a pleasure for me to acknowledge my mentors, Antonio Capolupo and Orlando Luongo, for their constant support. This work is based on the contribution to AstroMarche2 conference, September 2025, Unicam.

\section*{ORCID}

\noindent Francesco Lottatori - \url{https://orcid.org/0009-0002-6236-6895}

\end{document}